# Rational design and fabrication of versatile active colloidal molecules

Songbo Ni[1,2], Emanuele Marini[1,2], Ivo Buttinoni[1], Heiko Wolf [2]* and Lucio Isa[1]*

[1] Laboratory for Interfaces, Soft Matter, and Assembly, Department of Materials, ETH Zurich, Vladimir-Prelog-Weg 5, 8093 Zurich, Switzerland.

[2] IBM Research – Zurich, Säumerstrasse 4, 8803 Rüschlikon, Switzerland.

*corresponding authors: hwo@zurich.ibm.com, lucio.isa@mat.ethz.ch

**Abstract**

Active colloids, also known as artificial microswimmers, are self-propelled micro and nanoparticles that convert uniform sources of fuel (e.g. chemical) or uniform external driving fields (e.g. magnetic or electric) into directed motion by virtue of asymmetry in their shape or composition. These materials are currently attracting enormous scientific attention for two main end uses. First, active colloids are prototypical internally driven, out-of-equilibrium systems and their study has led to new emergent material properties, such as swarming and living crystallization. Secondly, they hold the promise to be used as micro- and nanoscale devices with tremendous potential from medical to environmental applications. However, the current fabrication of active colloids is limited in the programmability of materials, geometry, and modes of motion. Here, we use sequential capillarity-assisted particle assembly (sCAPA) to link microspheres of different materials into clusters of prescribed shapes ("colloidal molecules") that can actively translate, circulate and rotate powered by asymmetric electro-hydrodynamic flows. Further engineering of the geometry and composition provides active colloids that switch motion under external triggers or perform simple pick-up and transport tasks. Our fabrication strategy enables both physicists and engineers to design and create customized active colloids to explore novel fundamental phenomena in active matter and to investigate materials and propulsion schemes that are compatible with future applications.

**Introduction**

Active colloids, or artificial microswimmers, are microscale objects that can convert available, uniformly distributed energy into directional propulsion due to an intrinsic asymmetry in either their composition or geometry. Directed motion can be generated autonomously from the environment or can be imparted by external fields. Autonomous active colloids self-propel by exploiting surrounding chemical "fuels" (1–3), while externally activated colloids rather swim due to a local force induced by different sources, such as magnetic (4, 5), electric (6), light (7, 8), and ultrasonic fields (9, 10). As such, active colloids are a paradigmatic example of internally driven out-of-equilibrium thermodynamic systems, for which a broad range of emergent phenomena have been discovered, including chaining (11), clustering (12, 13), swarming (14), and phase separations (15). Moreover, they are central for the pursuit of miniaturized micro- and nanomachines, a vision that is beginning to take form with first proof-of-principle examples of autonomously powered microgears (16–18) and of vehicles for the delivery of drugs (19–21) or other cargoes (22), environmental remediation (23, 24), and lab-on-chip immunoassays (25).

To date, many different strategies have been used to fabricate symmetry-breaking active colloids (26). However, a method that allows their fabrication with full programmability in the choice of materials, composition and shape remains elusive. Such method would enable the rational design of artificial mircoswimmers with prescribed modes of motion and even functionality. This limitation is commonly recognized as one of the major hurdles preventing the further development of active colloids in a broad range of applications (27, 28). Here, we use sequential capillarity-assisted particle assembly (sCAPA) to fabricate a library of multi-material active colloidal molecules, i.e. clusters of microspheres assembled in prescribed geometries (29), whose swimming velocity and type of motion can be rationally designed. sCAPA is a recently developed method (29, 30) whose key feature is the ability to fully program the shape and the composition of hybrid colloidal molecules, here exploited for fabricating versatile microswimmers.

Actuation of asymmetric colloidal molecules can be realized in different ways (1–10). Here, we use electro-hydrodynamic flows to power active colloidal molecules moving over an electrode in a vertical AC field. The actuation mechanism was recently described based on a simple colloidal molecule (6), i.e. an asymmetric dimer. A charged dielectric particle near an electrode deforms the vertical electric field, which develops tangential components able to locally drive induced charges on the electrode and trigger electrohydrodynamic (EHD) flows (31). This EHD flow is highly dependent on surface chemistry, dielectric properties, surface charge and geometry of the colloidal particles, and is symmetric around spherical or homogeneous particles. However, introducing compositional or geometrical variations, e.g. in a heterogeneous dimer, breaks the symmetry, and unbalanced EHD flows drive a net propulsion. The direction and speed of this active motion are determined by the geometry and the degree of asymmetry generated by the properties of the individual components of the specific colloidal molecule. Furthermore, theoretical (31) and experimental (6) studies showed that the speed of the EHD flows, and thus the swimming speed of the active colloidal molecules, is proportional to the field strength squared, affording facile control of the propulsion velocity with an external control parameter.

In this work, we use the same driving mechanism to conceptually demonstrate that sCAPA is a general tool to fabricate customized active colloids. Using sCAPA to combine a wide variety of microspheres into clusters of prescribed shapes, we design and create active colloidal molecules out of different materials and with geometries that make them translate at various speeds, circulate, rotate, or switch between these modes of motion and even perform simple cargo pick-up and transport tasks in crowded environments.

**Results and discussions**

The fabrication, harvesting and actuation process of the colloidal molecules (CMs) is shown in Figure 1a. Briefly, microspheres from different materials are sequentially assembled in templates. The geometry of the template defines the shape of the CMs, and the assembly sequence determines their composition. The assembly of all CMs is detailed in Figure S1. After assembly and linking via sintering, CMs are dispersed in water and confined in an observation cell that consists of a bottom gold electrode, a top ITO electrode and a PDMS spacer with a thickness between 200 µm and 300 µm. AC electric fields with peak-to-peak voltages $V_{pp}$ between 4 and 30 V and frequencies from 500 Hz to 1.5 kHz are applied to actuate the CMs. The zoom in Figure 1a shows a sketch of the unbalanced EHD flows around a dumbbell-like CM. In this work, a library of complex active CMs is rationally fabricated, with shapes beyond dumbbells and with multiple material combinations. Examples of isolated CMs that translate and rotate are given in Figure 1b-f and Movie S1 – S5. They are constructed using amine-functionalized (green, 1.1 µm) and non-functionalized (red, 1 µm) polystyrene (PS) particles (see Materials and Methods for more details).

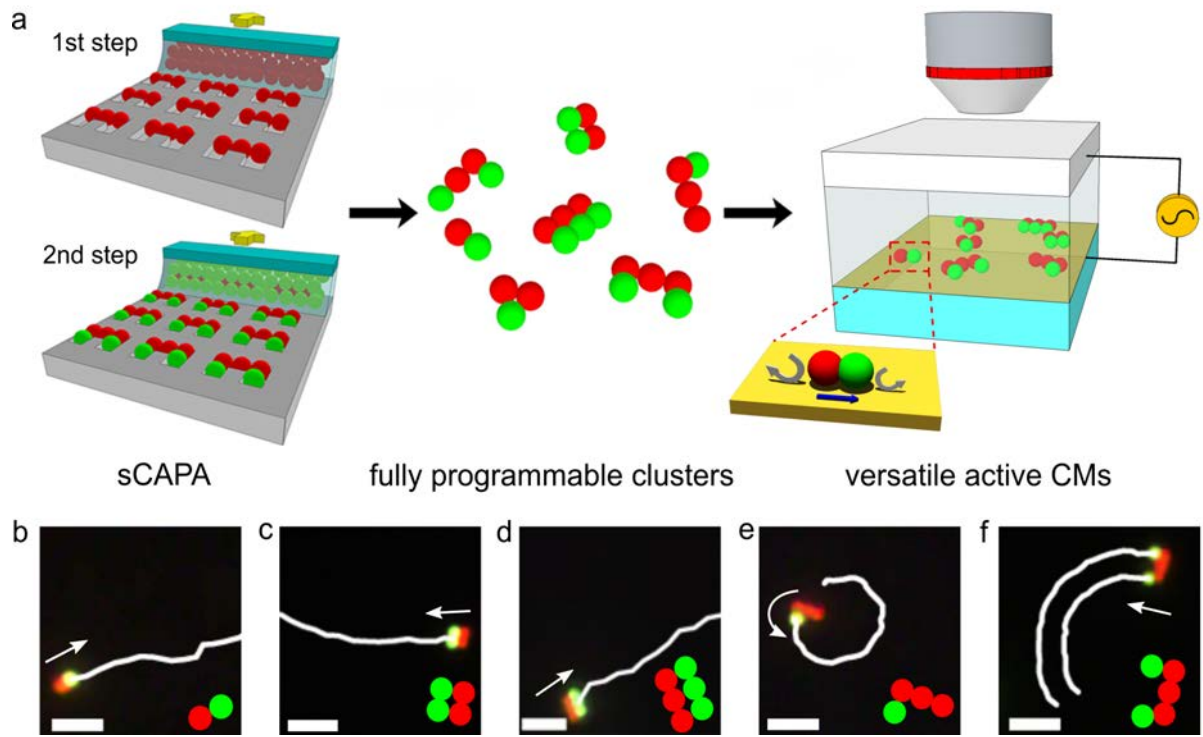

**Figure 1. Fabrication of versatile active colloidal molecules**. (a) Scheme of the process from assembly of CMs by sCAPA, to the harvesting and actuation in an AC electric field. The zoom schematically illustrates a possible unbalanced EHD flow around a dumbbell-like CM. (b – f) Fluorescence microscopy snapshots overlaid with the trajectories of active CMs of different geometries. The insets schematically show the details of the structures of the CMs. The arrows represent the directions of motion. The scale bars are 5 μm.

One of the main advantages of sCAPA is its flexibility in integrating different materials into a single CM, allowing a broad spectrum of combinations of functional properties. To demonstrate the impact that material choice has on the motion of active CMs, we have fabricated dumbbell-like microswimmers combining both organic PS particles (non-functionalized and amine-functionalized) and inorganic particles of $SiO_2$, $TiO_2$ and $SiO_2$ with embedded magnetic nanoparticles (henceforth referred to as $SiO_2$-Mag). Additional details are given in the SI, Figure S2. All of them exhibit self-propulsion, and their swimming speed $v$ and rotational diffusion time $\tau_R$ are calculated from the particle trajectories according to Bechinger et al. (27) and Wang et al. (32), i.e. respectively extracting the speed by fitting the short-time-scale mean square displacements and obtaining $\tau_R$ as a characteristic time scale for the decorrelation of the direction of the velocity vectors (see Materials and Methods).

For all these dumbbells, $v$ increases linearly with the square of the field strength (Figure S3), which is characteristic of propulsion from EHD flows (6, 31). We tune the swimming speed further by choosing different material combinations, while leaving the external field and the rotational diffusion time (Figure 2a inset) unaltered. Figure 2a shows the swimming speed of different dumbbell-like CMs, with the schemes indicating the propulsion direction. The variation in the swimming speed between the different types of dumbbells is attributed to the variations in the dielectric properties of the particles composing the lobes, i.e. their zeta potential $\zeta$ (see Materials and Methods) and Stern layer conductance $\sigma$. However, complete predictability of the swimming velocity and direction remains an open issue, mainly due to the challenging determination of the Stern layer conductance that is sensitive to the details of surface composition and its charging mechanism (6). On the other hand, the versatility of sCAPA in material choice enables the incorporation of functional particles in the active CMs, e.g., catalytic $TiO_2$ and magnetic $SiO_2$-Mag moieties, opening new avenues for the fabrication of new types of artificial microswimmers.

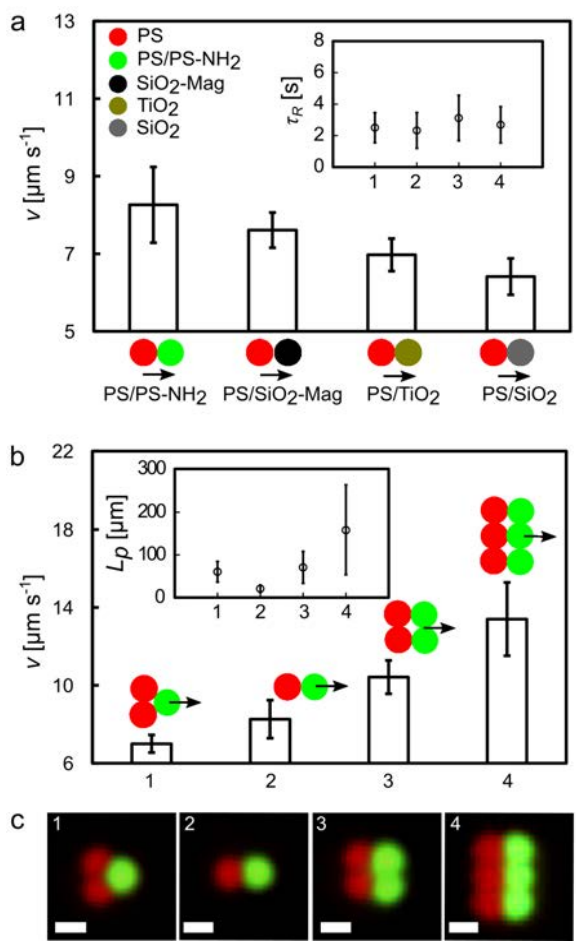

**Figure 2. Translating active colloidal molecules in different materials and shapes.** (a) Swimming velocities $v$ plotted for dumbbell-like active CMs in different material combinations, with an applied field of 32 V/mm at 1.3 kHz. Different materials are labeled in different colors as indicated in the legend. The schemes below the horizontal axis depict the corresponding dumbbells and their propulsion directions. The inset shows the rotational diffusion time $\tau_R$ for the active CMs in the same order. (b) Ranking of $v$ for active CMs of different shapes, with an applied field of 32 V/mm at 1.3 kHz. The arrows show the propulsion directions. The inset reports the persistence length $L_p$ of the active CMs. The fluorescence images in (c) show the corresponding active CMs labeled by the same numbers. The scale bars are 1 µm.

Full control on the colloidal molecules' shape is another key advantage of sCAPA. Using more complex assembly templates, we also demonstrate that simple variations in the CMs' geometry tremendously affect their ability to follow long straight paths, i.e., influence the persistency of their active trajectories. Figure 2b shows a range of different translating CMs (see Figure S2 and S4 for details of their structure) whose trajectories' persistence lengths $L_p = v \cdot \tau_R$ span more than two orders of magnitude. In contrast, $v$ and $L_p$ are decoupled by changing the architecture, so that a trimer and a four-pack have similar persistence lengths, but different propulsion velocities (CM type 1 and 3 in Figure 2b-c). Although the exact link between shape and propulsion speed requires a complete 3D simulation of the electro-hydrodynamic flows around each specific CM, which is out of the scope of the current manuscript, our findings indicate that the trajectories of our active CMs can be successfully engineered. We can design, for instance, CMs that move in straight lines over large distances for effective transport in long narrow channels or CMs that reorient frequently to promote effective local remediation or motion in tortuous porous media.

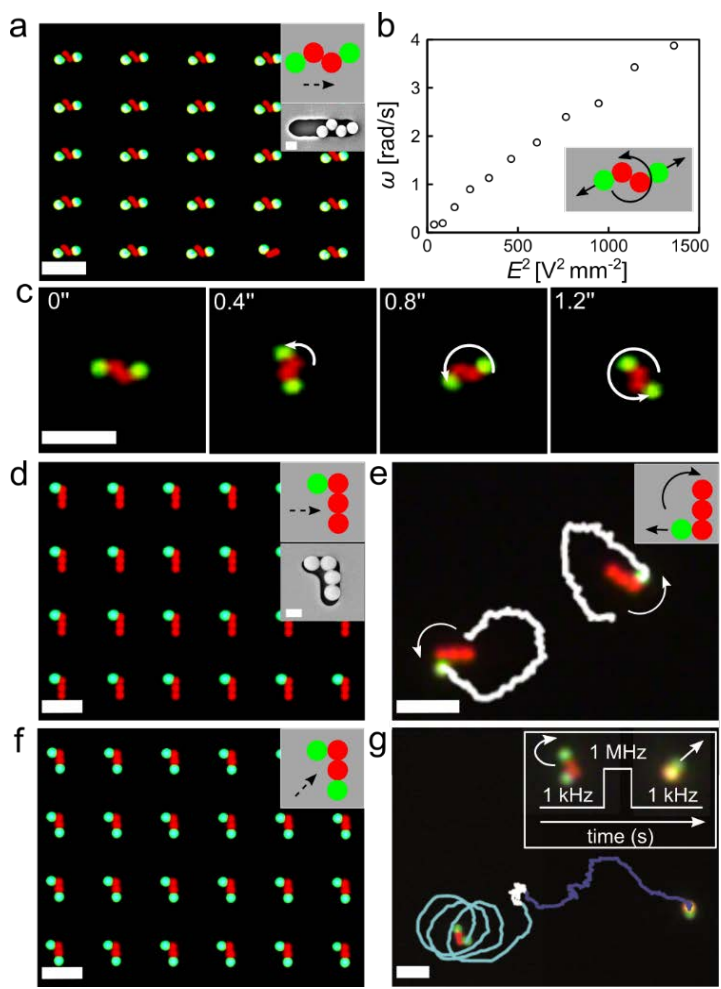

**Figure 3. Active colloidal molecules with versatile modes of motion: rotational, circular and reconfigurable motion.** (a - c) Rotating CMs. (a) Fluorescence image of colloidal rotors on the template (see Materials and Methods). The insets show the scheme of the structure of the CMs (the dashed arrow represents the assembly direction) and an SEM image of a colloidal rotor on the template. (b) The rotation speed ω plotted versus the square of the electrical field, $E^2$. ω increases linearly with $E^2$ as the speed $v$ does for the translating CMs. The inset schematically illustrates the rotation mechanism. (c) Snapshots of the CM in rotation at different times. (d – e) Circulating CMs. (d) Fluorescence image of L-shaped CMs on the template. The insets also show the corresponding scheme of the structure's assembly and an SEM image of the CM on the template. (e) Snapshot of L-shaped CMs during circular motion, overlaid with their trajectories. The solid arrows represent the directions of circular motion, whose mechanism is illustrated in the inset. (f - g) Reconfigurable CMs. (f) Fluorescence image of L-shaped active CMs with propulsion along two arms on the template. The inset again shows the CM scheme and the assembly direction (dashed arrow). (g) Trajectory showing the switch from circular (blue) to translational (purple) motion, overlaid with the initial and the last frame. The inset depicts the sequence of applied field frequency (E = 28 V/mm) and the corresponding switch in CM orientation. The scale bars are 5 μm in (a) – (g) and 1 μm in the insets of (a) and (d).

The ability to control the shape and composition of our active CMs enables rational design not only for tuning translational motion but also for enabling more complex modes of motions. Here, we fabricate colloidal rotors and L-shaped swimmers that can switch the mode of motion from circular to translational upon application of high-frequency pulses.

Figure 3a and its insets show a fluorescence image, an SEM image and the assembly scheme of colloidal rotors. The assembly is done in a trap that is slightly wider than the particle diameter (1.6 µm versus 1–1.1 µm), and sequential filling results in two active dumbbell units forming a zig-zag CM (see Figure S1 for details). The propulsion of the two dumbbell units in opposite directions, but with an offset, generates a torque and causes the CM to rotate (inset to Figure 3b). Remarkably we find that the CM's angular speed depends linearly on the square of the field strength (Figure 3b), analogous to the propulsion of dumbbells, implying facile external control in applications. Figure 3c shows snapshots of the rotation at different times (Movie S6). Figure 3d shows a fluorescence image, an SEM image, and the assembly scheme of L-shaped CMs (see Figure S1 for assembly details). Because of the propulsion along the short arms (i.e., by a dumbbell), a torque is established along with directed propulsion (33), and the activated CM moves circularly as shown in Figures 3e and 1e. The engineering of L-shaped CMs can be extended so that switching between different modes of motion becomes possible, and we have fabricated a second type of L-shaped CMs containing another green PS-NH$_2$ particle at the end of the long arm (see Figure S1 for assembly details), as shown in Figure 3f and its inset, leading to propulsion along both arms. Their successful fabrication relies on the precise control of assembly direction with respect to the geometry of the L-shaped traps (see Figure S1 for more details). In Figure 3g and Movie S7, we demonstrate that the second type of L-shaped CMs switches from circular to translational motion upon changing the frequency of the driving field to modify the CM's orientation relative to the substrate (see the inset in Figure 3g). Initially the CM lies flat on the substrate and exhibits circular motion at the standard frequency of 1 kHz. Increasing the AC frequency to 1 MHz will cause the CM to stop its propelling motion and to stand upright, aligning the long arm with the electric field, as previously observed for other anisotropic particles (34). When the frequency is switched back to 1 kHz, the CM has no time to reorient, and now translates with the short arm on the substrate. The possibility of propulsion along both arms ensures that the switch from circular to translational motion will also take place if the short arm aligns with the MHz field (see Movie S8), as opposed to the first type of L-shaped CM. The reverse switch from translational back to circular motion is achieved by turning off the field, waiting for the CMs to lie flat again and finally turning the field on at 1 kHz. This simple example already demonstrates the capability of appropriately designed multifunctional colloidal clusters to act as reconfigurable smart motors.

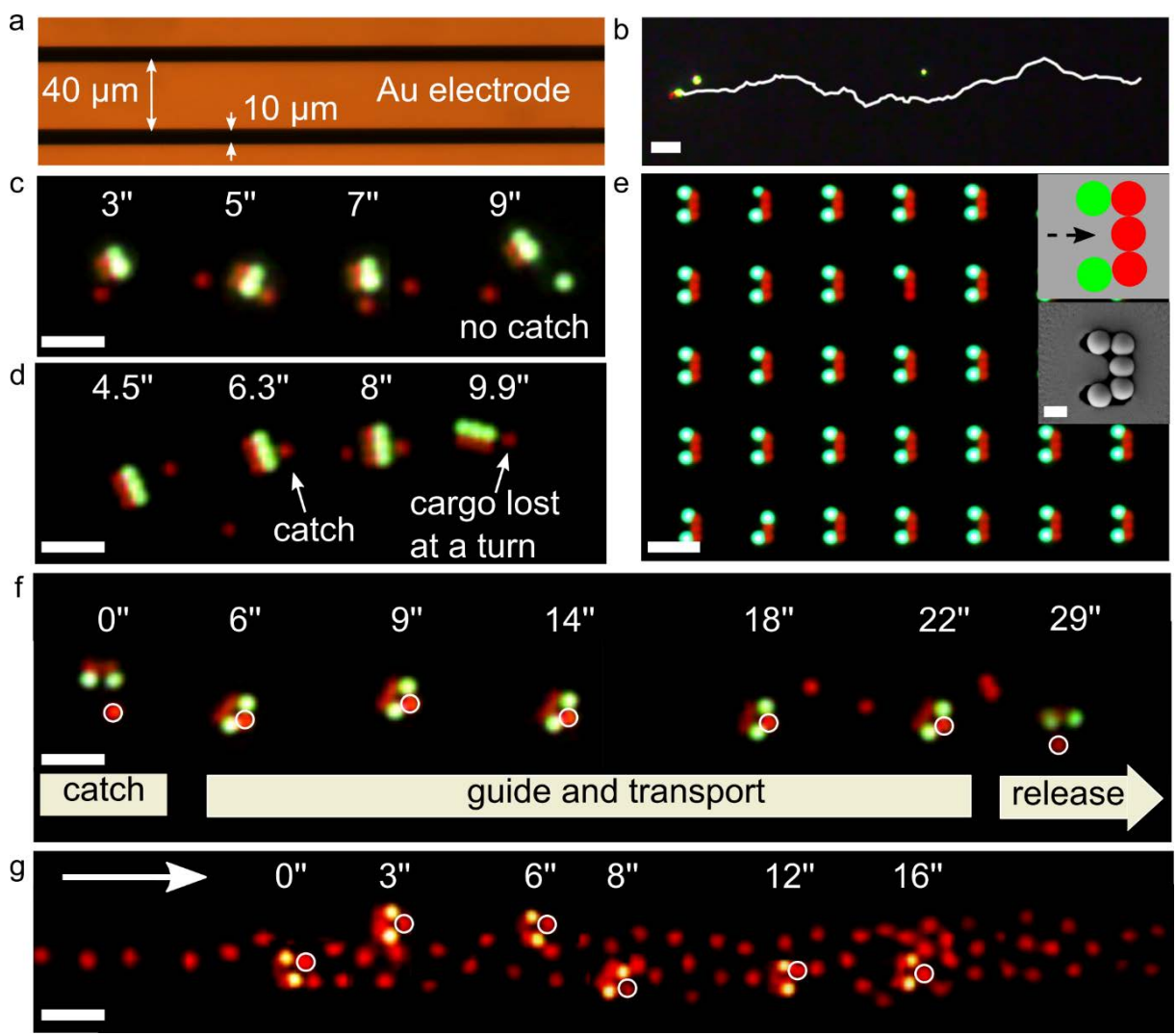

**Figure 4. Colloidal transporters: guiding, design and performance.** (a) Patterned electrode to guide active CMs. (b) Dumbbell-like CM traveling straight along the electrode. (c – d) Performance of square-like (c) and "six-pack" (d) colloidal transporters illustrated by a time series of fluorescence images. (e) Assembly of U-shaped CMs, with their scheme and SEM image in the inset. The dashed arrow represents the assembly direction. (f - g) U-shaped active CMs transport a cargo particle in an empty and in a crowded environment. The solid arrows represent the guiding and transport directions. Note that the actual distance traveled is larger than the image size because of the movement of the stage to follow the active CMs (see Movie S12 and S14). Scale bars are 5 µm (b – g) and 1 µm in the inset of (e).

Finally, simple specific tasks can be performed by rationally engineering the colloidal molecules. As a proof of concept, here, a specifically tailored CM is designed to "pick up", guide, transport, and release micro-objects. We performed "pick-up" tests for different CM shapes in a controlled environment defined by gold electrodes patterned by stripes (Figure 4a). The stripe generates an electro-hydrodynamic flow (35) that confines the CM within the pattern, but allows it to move along it (Figure 4b and Movie S9). This serves the purpose to confine the targets and the active CM to increase "pick-up" rates for the purpose of displaying the rationale behind their design. Figure 4b and Movie S9 show a dumbbell-like CM that travels in a straight line on the patterned electrode for a distance ($\approx$ 80 µm in Movie S9) much larger than its persistence length ($\approx$ 20 µm). We discuss three different designs of CMs that are potentially able to pick up and transfer a micro-object: a square (two dumbbells in parallel, Figure S4), a "six-pack" (three dumbbells in parallel, Figure S4) and a U-shaped CM. Single PS and PS-NH$_2$ spheres are intentionally dispersed to serve as cargoes. Movie S10 and Figure 4c show the square CM in action. The small cross section perpendicular to the moving direction leads to failure to pick up any particles. An improvement is seen by increasing the cross section using a "six-pack". Movie S11 and Figure 4d reveal that a "six-pack" CM can pick up a target particle, but that it loses it when changing direction. To protect the cargo during transport and avoid its loss during changes of direction, we designed and fabricated U-shaped microswimmers as shown in Figure 4e (for assembly details, see Figure S1). Figure 4f shows a time series demonstrating that these U-shaped CMs effectively pick up and transport a cargo along the patterned electrode stripe when the field is on and release it when the field is turned off (Movie S12). The effectiveness of this design is also demonstrated by showing that different types of target particles can be captured (PS-NH$_2$, Figure S6 and Movie S13) and transported, even in crowded environments (Figure 4g and Movie S14). The only failure mode observed in crowded environments is the exchange of the particle being transported with another one, as shown in Figure S7 and Movie S15. This proof of concept shows how the design of CMs can be used to define and optimize their function, and forms the basis for future implementations in which the geometry (longer side arms) or the material composition (magnetic beads for remote steering) may be tailored for specific applications.

**Conclusions**

Our results demonstrate that sCAPA enables an unprecedented level of engineering of a variety of colloidal molecules that actively move in vertical AC electric fields. The CMs are prepared from different materials and in various shapes. They can be tailored and designed to translate, rotate, switch the type of motion, and even pick up and transport cargoes. Given the flexibility that sCAPA provides in rationally designing and fabricating multifunctional clusters of colloidal particles, one can easily envision the fabrication of CMs powered by mechanisms other than electro-hydrodynamic flows or even by a combination of mechanisms. Future developments include scaling-up of particle production, downsizing of the assembled CMs towards the nanoscale and the development of predictive power to link the geometry, composition and propulsion strategies of the CMs to their active behavior. The challenges of scaling up fabrication increase as the structure of colloidal molecules becomes more complex. While the current throughput of dimers can reach a few millions per hour with a nearly 100% uniformity, more complex colloidal molecules require future process optimization (e.g. higher fidelity of the template and engineering of the meniscus shape) to increase the uniformity and thus the throughput. The size of the assembled particles can be downscaled (36) and currently, sCAPA of nanoparticles is being performed to test the limits of the method. Finally, increasing the complexity of CMs in the future also calls for in-depth fundamental studies to develop predictive power in the design of specific CMs with tailored motion and function, which may constitute the first steps toward the realization of colloidal micro- and nanomachines.

In conclusion, we envision sCAPA to become a powerful toolbox for the parallel fabrication of customized active colloids. Engineering of particle geometry and composition will help to identify new propulsion schemes, to screen and test the compatibility of the active colloids towards practical micro- and nanodevices and may contribute to unravel the complexity of the active systems ubiquitously found in nature.

**Materials and Methods**

**Colloidal suspensions**

Fluorescent non-functionalized PS particles with a diameter of 1 µm (red, R0100, polydispersity < 5%) and plain silica particles with a diameter of 1 µm (8100; polydispersity < 5%) were purchased from Thermo Scientific. Fluorescent amine-functionalized PS particles with a diameter of 1.1 µm (green, F-8765, polydispersity < 5%) were purchased from Life Technologies. Magnetic particles (iron-oxide nanoparticles in a silica matrix) with a diameter of 0.96 ± 0.05 µm ($SiO_2$-MAG-AR359) were purchased from microParticles GmbH. Titania ($TiO_2$) particles with a diameter of 1 µm (C-TIO-1.0, polydispersity < 5%) were purchased from Corpuscolar Inc. The zeta potential for all particles was measured by Zetasizer Nano (Malvern) and can be found in Figure S8.

**Fabrication of active colloidal molecules**

The fabrication of the assembly template in PDMS (poly(dimethylsiloxane), the sequential capillarity-assisted particle assembly, the linking, printing and harvesting of the CMs are all done according to the protocol reported in ref. (29).

**Actuation by a vertical AC field**

Aqueous suspensions of colloidal molecules fabricated by sequential capillarity-assisted particle assembly were placed in a cell consisting of a bottom gold electrode (100 nm of gold evaporated on a glass slide), a top ITO electrode and a PDMS spacer with a thickness between 200 µm and 300 µm. The colloidal molecules were actuated by AC fields with peak-to-peak voltages $V_{pp}$ between 4 and 30 V and frequencies from 500 Hz to 1.5 kHz (AC function generator, Agilent 33250A).

Switching of the L-shape colloidal molecule's motion modality was performed with an applied AC electrical field of 20 $V_{pp}$ and a spacer of 230 µm. The motion of the CMs was switched and controlled by changing the frequency from 1 kHz to 1 MHz and back to 1 kHz.

Guiding of CMs was performed on patterned gold electrodes. The electrodes were prepared by optical lithography and metal deposition (e-beam physical-vapor deposition, 100 nm Au) followed by a lift-off step.

**Imaging and analysis**

Colloidal molecules assembled on the template or printed on a substrate were imaged by bright-field, dark-field, and fluorescence optical microscopy (single-filer or dual-filter, Zeiss Axioscope) at various magnifications and by SEM (Leo 1550 Gemini, Carl Zeiss AG). ImageJ (W. S. Rasband, National Institutes of Health, Bethesda, MD) was used to compose overlays of fluorescence and bright-field images by combining different channels. All movies were grabbed at a frame rate of 7 or 10 fps using a dual filter in fluorescence optical microscopy, or using bright-field microscopy and a CCD camera. For aesthetic reasons, frames in Figure 3g, 4b-d, 4f-g, S6, and S7 were rotated by 90° as compared to the original movies.

The positions of the active colloidal molecules were acquired by tracking one particle using either an ImageJ plugin (37), customized software written in Matlab (MathWorks) or IDL (Harris Geospatial). The coordinates of the tracked particle in each frame were used to calculate the swimming speed according to ref. (27). The mean square displacement (msd) $\langle [\Delta L(t)]^2 \rangle$ was plotted against time *t*, which can be expressed by the following equation (27):

$$\langle [\Delta L(t)]^2 \rangle = (4D_0 + 2v^2\tau_R)t + 2v^2\tau_R^2(e^{-\frac{t}{\tau_R}} - 1), \quad (1)$$

and for times close to the rotational diffusion time $\tau_R$, the msd curve can be fitted by the following equation after a second-order expansion:

$$\langle[\Delta L(t)]^2\rangle = 4\, D_0 t + v^2 t^2, \quad (2)$$

where $D_0$ is the 2D translational diffusion coefficient of a passive colloidal molecule and $v$ is its swimming speed. $D_0$ can be measured from the slope of the MSD of the passive colloidal molecules without any external field applied. The translational diffusion coefficients of a passive dumbbell, a passive triangle, a passive square and a passive "six-pack" colloidal molecule were measured to be 0.179 μm²/s, 0.148 μm²/s, 0.097 μm²/s, and 0.058 μm²/s, respectively.

The rotational diffusion time $\tau_R$ was estimated according to ref. (32), based on the analysis of the time correlation between the directions of the swimming speeds from an initial time $t_0$ and after a time lag $t$, $V(t_0)$, and $V(t)$, respectively. As the swimming steps are much larger than the Brownian fluctuations, $\tau_R$ can be extracted using the following fitting for the correlation curve:

$$\langle \underline{V}(t) \cdot \underline{V}(t_0)\rangle = v^2 e^{-t/\tau_R} \quad (3)$$

**Supplementary Information:**

The Supplementary Information contains additional figures and movies.

**Acknowledgments:**

We thank U. Drechsler, R. Stutz and S. Reidt for help with the mask and master fabrication, A. Knoll, U. Duerig, K. Carroll, H. Löwen and B. ten Hagen for discussions and advice, C. Bolliger for assistance with the manuscript, and R. Allenspach and W. Riess for continuous support. L.I., S.N., and I.B. acknowledge financial support from the Swiss National Science Foundation (grant PP00P2_144646/1).

**Author Contributions:**

SN, IB, HW and LI conceived and designed the study. SN and EM performed the experiments. All authors analyzed the data. SN, IB, HW and LI wrote the manuscript. All authors discussed the results and commented on the manuscript.

# Supplementary Information

*Propulsion mechanism by electrohydrodynamic (EHD) flow*

A highly charged dielectric particle can deform the local electrical field in the vicinity of a conducting electrode. The tangential component of the deformed electric field can, as a consequence, locally drive the induced charges on the electrode and cause an electrohydrodynamic (EHD) flow in an alternating potential. These EHD flows have been previously studied from a fundamental perspective[1] and applied for colloidal assembly[2,3]. More recently, breaking the symmetry of the EHD flow using asymmetric colloidal molecules has also been investigated by Wu and coworkers for locomotion[4–6]. This mechanism is different from the well-known propulsion mechanism driven by induced-charge electro-osmosis (ICEO) where, usually, a metallic patch has to be incorporated in an asymmetric particle[7], although both propulsions are activated in an AC electric field. The inset to Figure S3 schematically illustrates the unbalanced EHD flows for the example of a PS-SiO$_2$ dimer. The actual directions of the EHD flows at both sides of PS-SiO$_2$ dimers have been visualized experimentally[4] and the unbalanced EHD flow can propel the dimer with the SiO$_2$ lobe moving in the front. Here, we fabricate PS-SiO$_2$ dimers by sCAPA and find the same swimming direction and the same scaling law with the electric field strength (Figure S3). The swimming speed *v* increases linearly with the square of the field strength E$^2$, which pinpoints the origin of the driving mechanism on EHD flows, as described by other experimental and modeling studies[1,4]. In particular, the velocity of the electro-hydrodynamic flow is $U_{EHD} \sim C \frac{E^2 \varepsilon \varepsilon_0 \kappa H}{\mu} \frac{K' + \bar{\omega} K''}{1 + \bar{\omega}^2}$, where *C* is a geometrical factor (e.g. from particle size and shape), *E* is the applied electrical field strength, $\varepsilon\varepsilon_0$ is the solvent permittivity, $\kappa^{-1}$ is the Debye length, *H* is the separation between two electrodes, $\mu$ is the viscosity of the medium, $\bar{\omega}$ is the dimensionless frequency ($\bar{\omega} = \omega H / \kappa D$, where *D* is the ion diffusivity), *K'* and *K''* are real part and imaginary part of the complex polarization coefficient[8], respectively. Both *K'* and *K''* are sensitive to the field frequency $\omega$, the particle radius *r*, the ion diffusivity *D*, the Debye length $\kappa^{-1}$ and the particle surface chemistry, i.e. the zeta potential $\zeta$ and the Stern layer conductance $\sigma$. This implies that, any asymmetry in zeta potential, Stern layer conductance (both often related to surface composition) and size can lead to an unbalanced EHD flow around the lobes of a colloidal molecule, which further drives net propulsion.

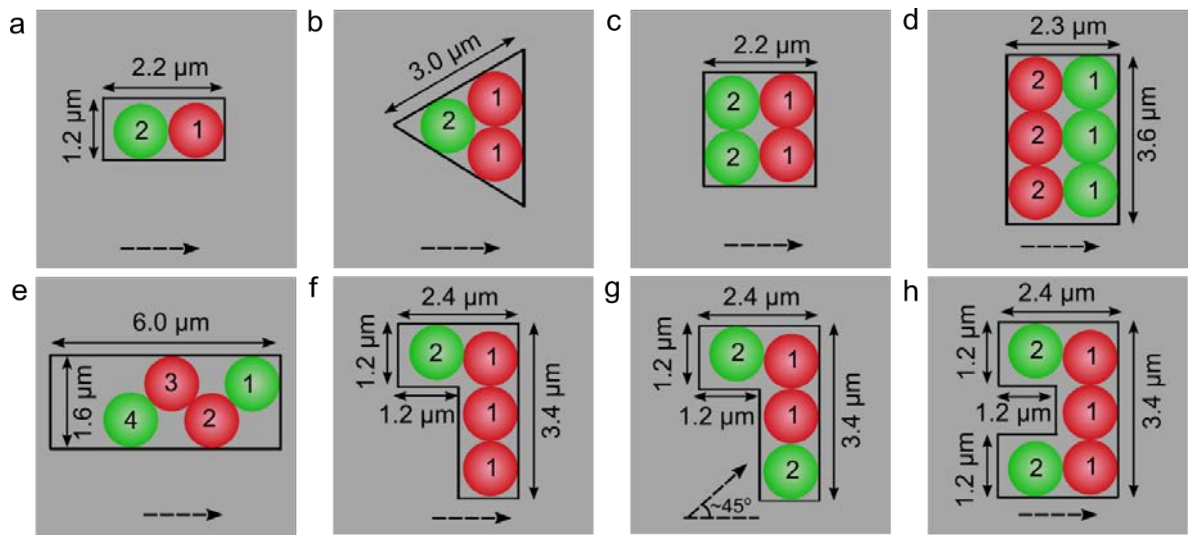

**Figure S1. Design and assembly of different active colloidal molecules.** (a – h) Schemes showing the lateral dimensions of the traps used, the assembly directions and sequences. All traps have a depth varying between 500 and 550 nm. All particles have a diameter of 1 to 1.1 µm. Dashed arrows indicate the assembly directions. The numbers on the spheres represent the sequence in the assembly, e.g., 1 stands for the first step and 2 for the second step, etc. Spheres in different color stand for different materials. (a – d) Colloidal molecules with different shapes: dumbbells (a), triangles (b), squares (c) and "six-packs" (d) can be assembled sequentially in two steps. (e) A zig-zag colloidal chain can be assembled in four steps by controlling the width of the trap. Tighter traps (trap width close to particle diameter) lead to straighter chains, whereas wider traps lead to colloidal chains with a zig-zag configuration. (f – g) Two types of L-shaped colloidal molecules. (f) The first type can be assembled in two steps, with the assembly direction perpendicular to the long arm of the CM. (g) The second type can also be assembled in two steps, but the assembly direction is turned by 45°, as shown by the dashed arrow. (h) U-shaped colloidal molecules can be assembled in two steps using the trap design shown here.

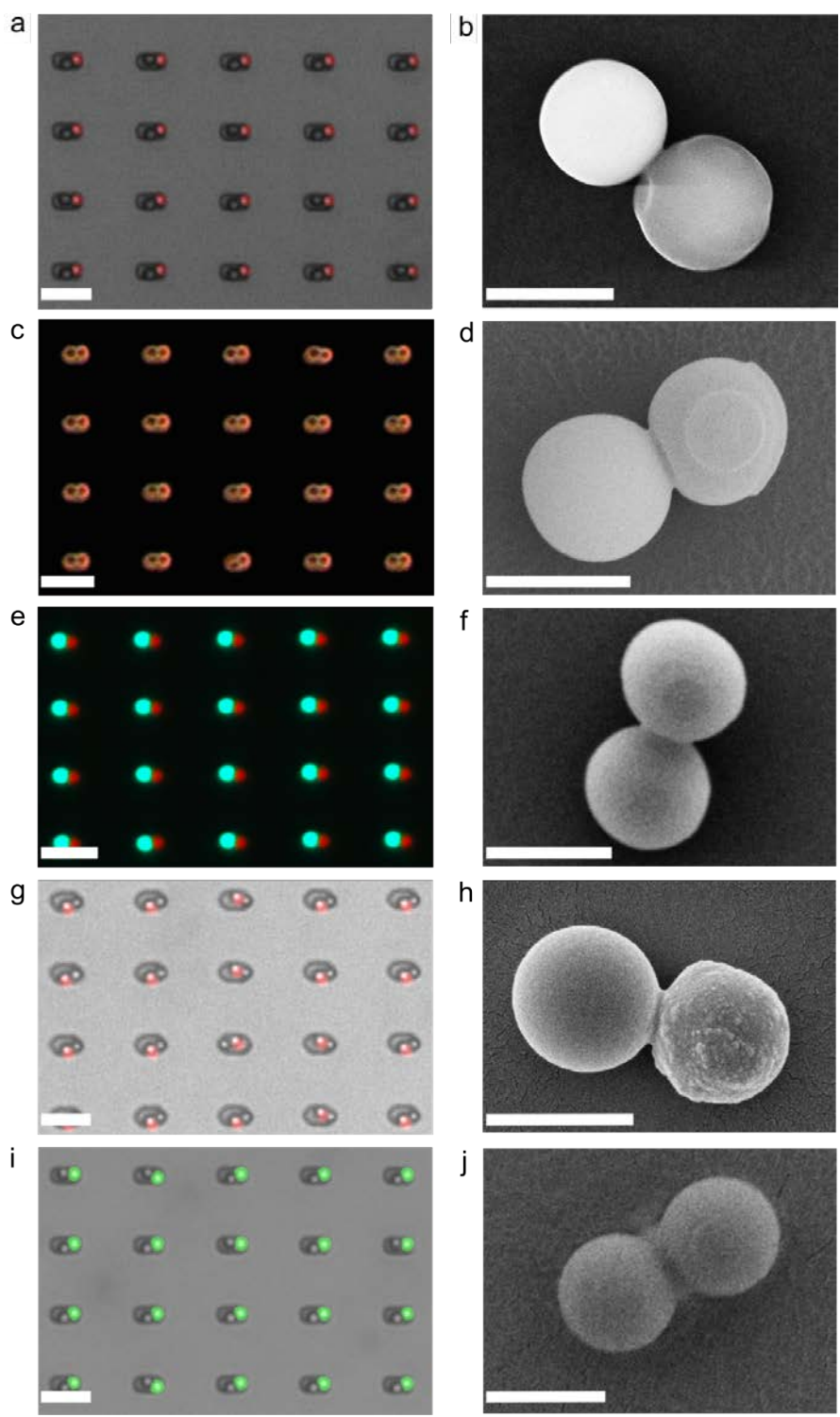

**Figure S2. Optical and SEM images of all the dumbbell-like colloidal molecules.** (a-b) Overlay of a fluorescence and a bright-field image (a) and SEM image after printing to a silicon substrate (b) for PS/SiO$_2$ dumbbell. (c-d) Dark-field image (c) and SEM image after printing to a silicon substrate (d) for PS/TiO$_2$ dumbbell. (e-f) Dual-channel fluorescence image (e) and SEM image after printing to a silicon substrate (f) for PS/PS-NH$_2$ dumbbell. (g-h) Overlay of a fluorescence and a bright-field image (g) and SEM image after printing to a silicon substrate (h) for PS/SiO$_2$-Mag dumbbell. (i-j) Overlay of a fluorescence and a bright-field image (i) and SEM image after printing to a silicon substrate (j) for PS-NH$_2$/SiO$_2$ dumbbell. Scale bars: 5 μm (a, c, e, g, i) and 1 μm (b, d, f, h, j).

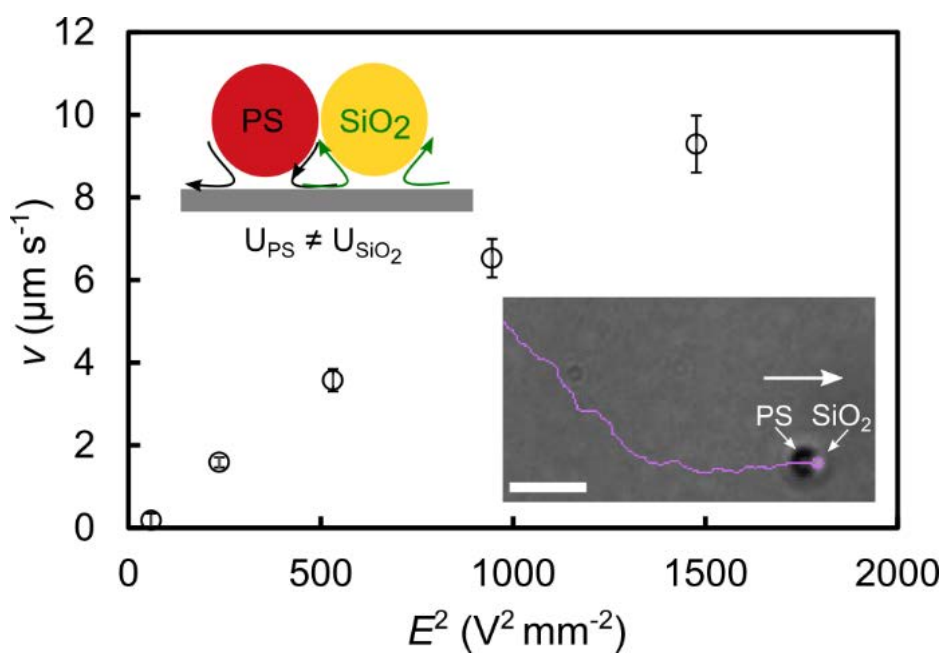

**Figure S3. Swimming speed *v* plotted versus the square of the electrical field strength $E^2$ for PS/SiO$_2$ dumbbells.** *v* increases linearly with $E^2$. The bright-field image (inset) shows a PS/SiO$_2$ dumbbell with its trajectory. The scheme indicates the propulsion direction. Scale bar: 5 µm.

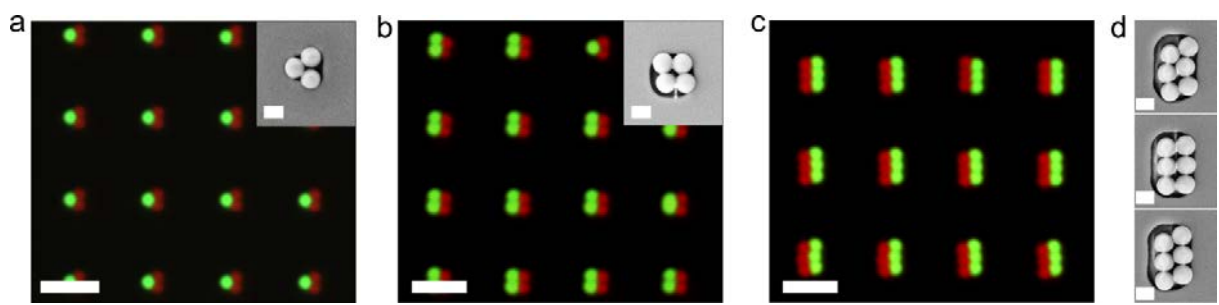

**Figure S4. Fluorescence and SEM images for triangle (a), square (b) and six-pack-like (c) colloidal molecules.** (d) The structure of six-pack-like CMs can vary depending on the trap used, which accounts for the larger error bar present in Figure 2b in the main text. Scale bars: 5 µm (a, b, c), 1 µm in the insets of (a, b) and in (d).

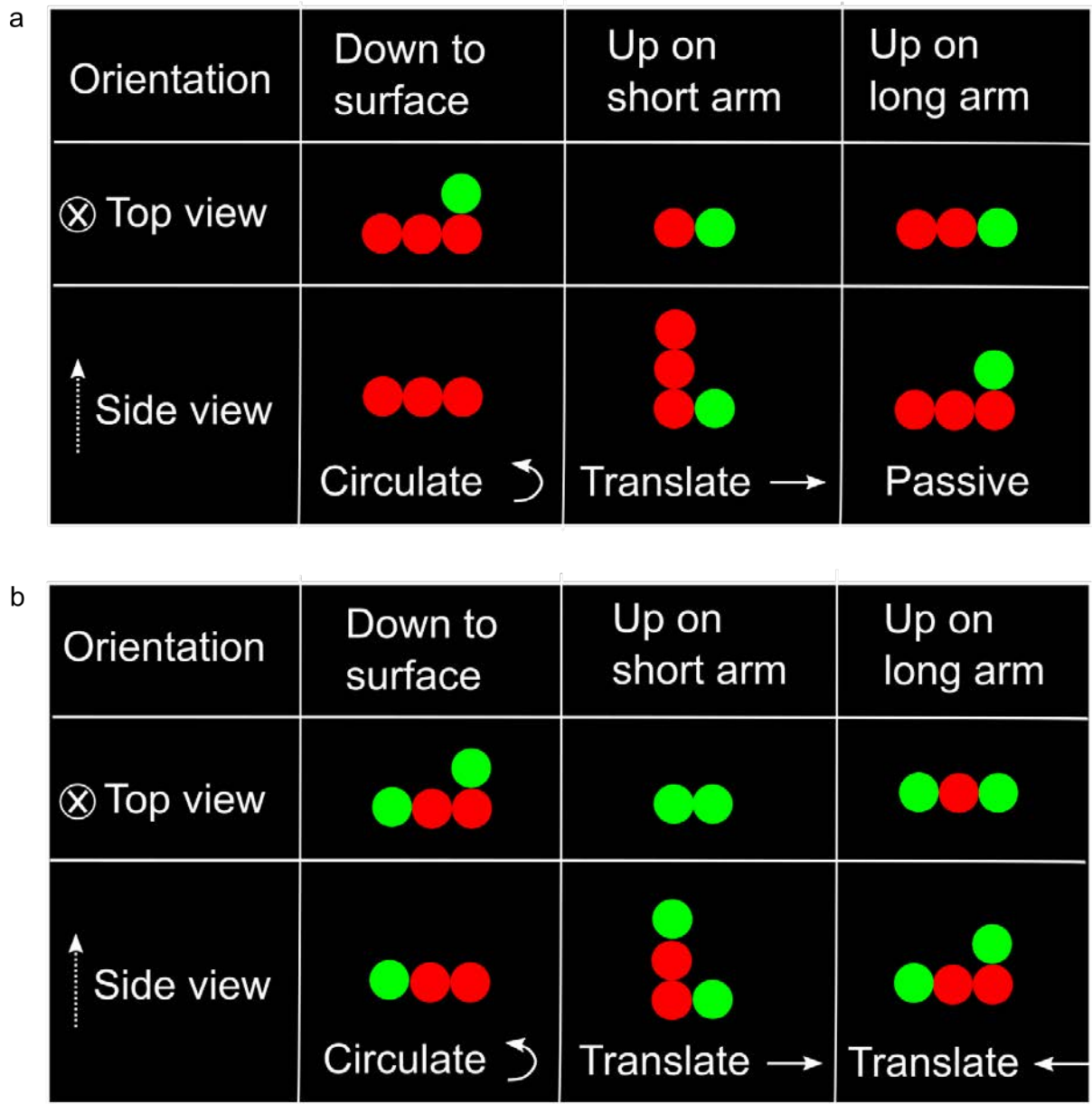

**Figure S5. Comparison of the types of motion for two types of L-shaped active colloidal molecules** at different orientations, without (a) and with (b) a green particle at the end of the long arm (red: PS particles; green: PS-NH$_2$ particles). (a) L-shaped active colloidal molecules without a green particle at the long arm perform circular motion while lying flat and translate while standing up on the short arm, but do not exhibit any active motion while standing up on the long arm because there is no asymmetric electro-osmotic flow parallel to the substrate. (b) L-shaped active colloidal molecules with a green particle also perform circular motion while lying flat, but translate while standing up, both on the short and on the long arm.

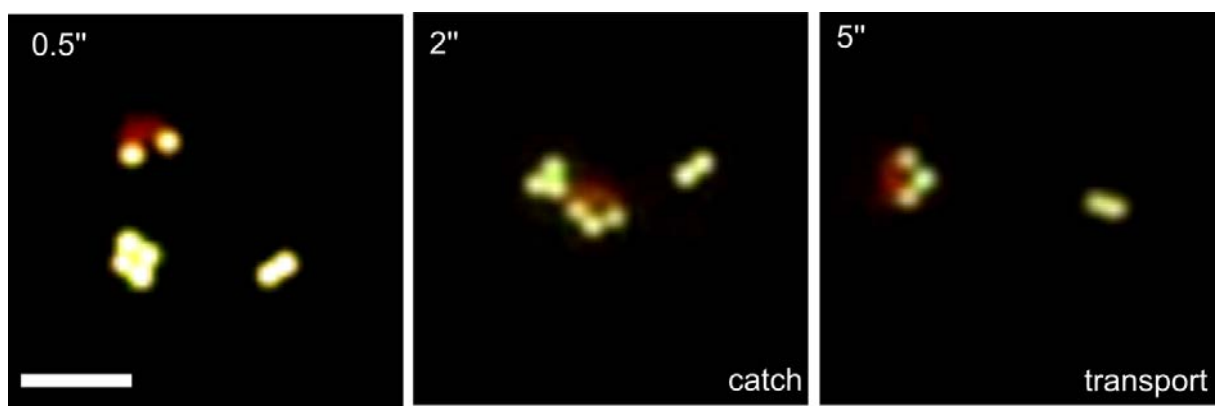

**Figure S6. Transport of a PS-NH$_2$ target particle by the U-shaped active colloidal molecules on a patterned electrode.** A U-shaped CM picks up a PS-NH$_2$ (green) particle and transports it, as shown by the time series of fluorescence images (corresponding Movie S13). Scale bar: 5 µm.

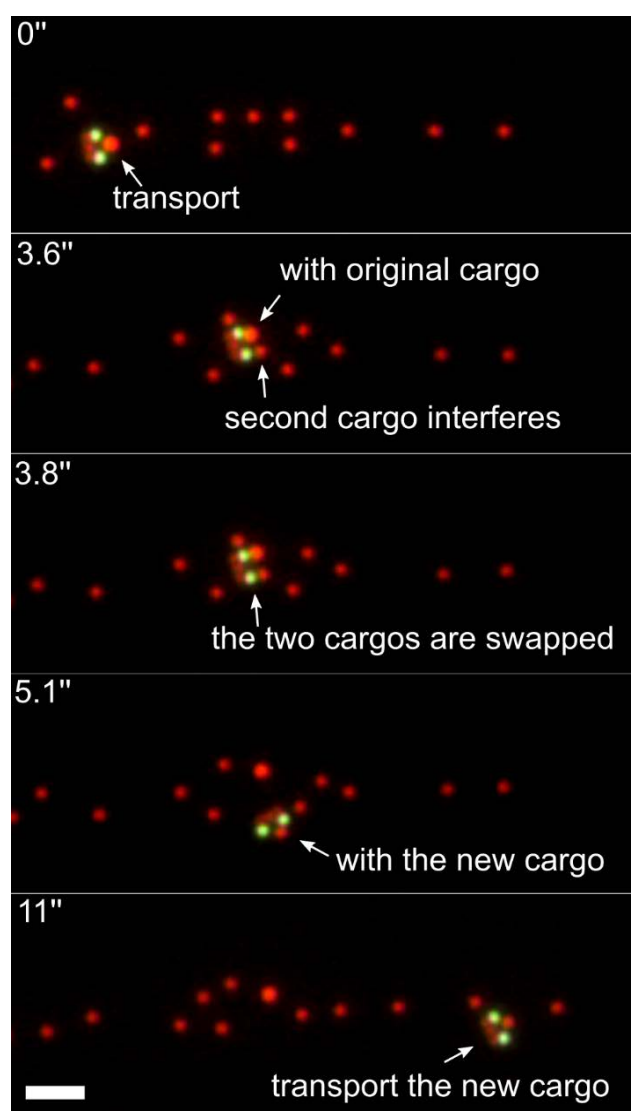

**Figure S7. Cargo swap by a U-shaped active colloidal molecule in a crowded environment.** Initially, the U-shaped CM is transporting its cargo particle through a crowded environment of other PS

particles. At the time of 3.6'', another particle (red, PS) is close enough to hinder the motion of the U-shaped CM, in particular the motion on one arm. This disturbance leads to the situation in which the cargo and the second particle are in close proximity to the CM and become equally probable to be picked up (see the time at 3.8''). In the example selected, the cargo is swapped, and the U-shaped CM travels on with the new cargo particle (see frames at 5.1" and 11"). Scale bar: 5 µm.

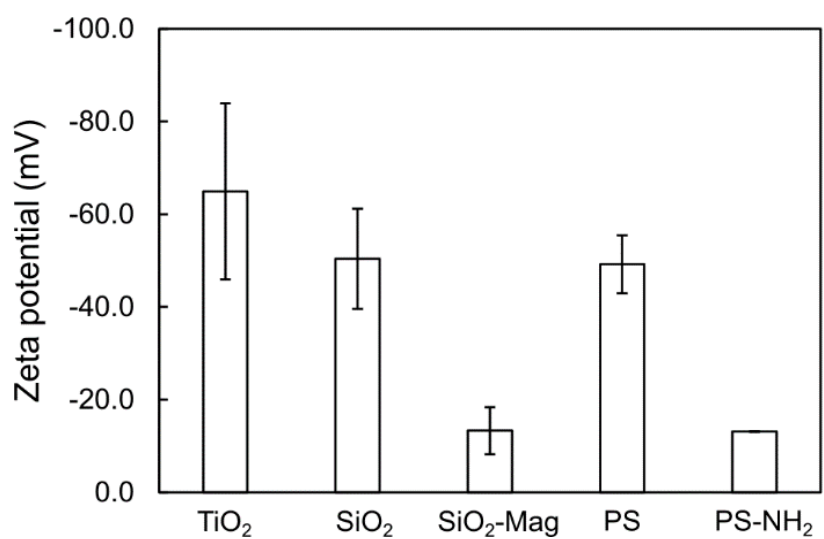

**Figure S8. Zeta potential measured for different single particles used in this study.**

SUPPLEMENTARY MOVIES:

Movie S1. An actuated and self-propelling, dumbbell-like colloidal molecule. The movie was captured by dual-channel fluorescence microscopy at 10 fps, with an applied AC electrical field at 1.3 kHz, 25 V (peak to peak voltage) and a spacer of 220 µm.

Movie S2. An actuated and self-propelling, square-like colloidal molecule. The movie was captured by dual-channel fluorescence microscopy at 10 fps, with an applied AC electrical field at 1.3 kHz, 25 V (peak to peak voltage) and a spacer of 220 µm.

Movie S3. An actuated and self-propelling, six-pack-like colloidal molecule. The movie was captured by dual-channel fluorescence microscopy at 10 fps, with an applied AC electrical field at 1.3 kHz, 22 V (peak to peak voltage) and a spacer of 245 µm.

Movie S4. An actuated and self-propelling, L-shaped colloidal molecule. The movie was captured by dual-channel fluorescence microscopy at 10 fps, with an applied AC electrical field at 1kHz, 26 V (peak to peak voltage) and a spacer of 250µm.

Movie S5. An actuated and self-propelling, U-shaped colloidal molecule. The movie was captured by dual-channel fluorescence microscopy at 10 fps, with an applied AC electrical field at 1 kHz, 20 V (peak to peak voltage) and a spacer of 240 µm.

Movie S6. An actuated and rotating, zigzag-like colloidal molecule. The movie was captured by dual-channel fluorescence microscopy at 7 fps, with an applied AC electrical field at 1 kHz, 24 V (peak to peak voltage) and a spacer of 230 μm.

Movie S7. A reconfigurable, L-shaped colloidal molecule, switching from circular to translational (standing on the short arm) motion. The movie was captured by dual-channel fluorescence microscopy at 10 fps, with an applied AC electrical field (1 kHz to 1 MHz back to 1 kHz), 20 V (peak to peak voltage) and a spacer of 230 μm.

Movie S8. A reconfigurable, L-shaped colloidal molecule, switching from circular to translational (standing on the long arm) motion. The movie was captured by dual-channel fluorescence microscopy at 10 fps, with an applied AC electrical field (1 kHz to 1 MHz back to 1 kHz), 20 V (peak to peak voltage) and a spacer of 230 μm.

Movie S9. An actuated, dumbbell-like colloidal molecule swimming along the straight patterned electrode. The movie was captured by dual-channel fluorescence microscopy at 10 fps, with an applied AC electrical field at 1 kHz, 20 V (peak to peak voltage) and a spacer of 230 μm.

Movie S10. A failed pick up by a square-like active colloidal molecule. The movie was captured by dual-channel fluorescence microscopy at 10 fps, with an applied AC electrical field at 1 kHz, 20 V (peak to peak voltage) and a spacer of 300 μm.

Movie S11. A failed pick up and transport by a six-pack-like active colloidal molecule. The movie was captured by dual-channel fluorescence microscopy at 10 fps, with an applied AC electrical field at 1 kHz, 20 V (peak to peak voltage) and a spacer of 300 μm.

Movie S12. A complete and successful pick up (red particle), transport and release by a U-shaped active colloidal molecule. The movie was captured by dual-channel fluorescence microscopy at 10 fps, with an applied AC electrical field at 1 kHz, 20 V (peak to peak voltage) and a spacer of 300 μm.

Movie S13. A successful pick up (green particle) and transport by a U-shaped active colloidal molecule. The movie was captured by dual-channel fluorescence microscopy at 10 fps, with an applied AC electrical field at 1 kHz, 20 V (peak to peak voltage) and a spacer of 290 μm.

Movie S14. A U-shaped active colloidal molecule swimming in a crowded environment after picking up a cargo. The movie was captured by dual-channel fluorescence microscopy at 10 fps, with an applied AC electrical field at 1 kHz, 20 V (peak to peak voltage) and a spacer of 290 μm.

Movie S15. A U-shaped active colloidal molecule swapping cargo during transportation. The movie was captured by dual-channel fluorescence microscopy at 10 fps, with an applied AC electrical field at 1 kHz, 20 V (peak to peak voltage) and a spacer of 300 μm.